\documentclass[preprint,showpacs,preprintnumbers,amsmath,amssymb]{revtex4}
\usepackage{graphicx}
\usepackage{dcolumn}
\usepackage{bm}
\usepackage{natbib}
\usepackage{amsthm}

\begin{document}

\preprint{}

\title{Irreversible Thermodynamics in Multiscale Stochastic Dynamical Systems}

\author{Mois\'es Santill\'{a}n}
\email{msantillan@cinvestav.mx}
\affiliation{Centro de Investigaci\'{o}n y Estudios Avanzados del IPN, Unidad Monterrey, Parque de Investigaci\'{o}n e Innovaci\'{o}n Tecnol\'{o}gica, 66600 Apodaca NL, M\'{E}XICO}

\author{Hong Qian}
\email{qian@amath.washington.edu}
\affiliation{Department of Applied Mathematics, University of Washington, Box 352420, Seattle, WA 98195,
USA}

\begin{abstract}
This work extends the results of the recently developed theory of a rather complete thermodynamic formalism  for discrete-state, continuous-time Markov processes with and without detailed balance. We aim at investigating the question that whether and how the thermodynamic structure is invariant in a multiscale stochastic system. That is, whether the relations between thermodynamic functions of state and process variables remain unchanged when the system is viewed at different time scales and resolutions. Our results show that the dynamics on a fast time scale contribute an \emph{entropic} term to the ``internal energy function'', $u_S(x)$, for the slow dynamics. Based on the conditional free energy $u_S(x)$, one can then treat the slow dynamics as if the fast dynamics is nonexistent. Furthermore, we show that the free energy, which characterizes the \emph{spontaneous organization} in a system without detailed balance, is invariant with or without the fast dynamics: The fast dynamics is assumed to reach stationarity instantaneously on the slow time scale; they have no effect on the system's free energy. The same can not be said for the entropy and the internal energy, both of which contain the same contribution from the fast dynamics. We also investigate the consequences of time-scale separation in connection to the concepts of quasi-stationaryty and steady-adiabaticity introduced in the phenomenological steady-state thermodynamics.
\end{abstract}

%\pacs{}

\maketitle

\section{Introduction}

Stochastic dynamics that can be described by a Markov process embody a rich thermodynamic structure. Recently, inspired by the discovery of the fluctuation theorem \citep{Evans:1993fk, Evans:1994uq,Gallavotti:1995,kurchan:1998,lebowitz,crooks,Wang:2002kx}, there is a growing interesting in concepts such as Gibbs entropy and free energy associated with Markov processes \citep{Qian:2002ys,Ge:2009,Ge:2010fk,esposito_prl:2010,esposito_pre1:2010,esposito_pre2:2010}. The free energy of a stochastic dynamics is intimately related to the relative entropy which has found great importance in the general theory of dynamical 
systems \cite{Lebowitz:1955,Qian:2001,Mackey:2003zr}.

In the very recent paper \cite{Ge:2010fk}, a rather complete thermodynamics has been presented for discrete-state, continuous-time stochastic Markov systems with or without detailed balance. The thermodynamics are characterized by: 
\begin{enumerate}
\item[(i)] A balance equation for the Gibbs entropy that includes a non-negative entropy production rate $\sigma$. 
\item[(ii)] A decreasing free energy $dF/dt\le 0$.
\item[(iii)] A decomposion of $\sigma$ into $-dF/dt$ and the house keeping heat $Q_{hk}$;
both being non-negative. 
\end{enumerate}
Assertion (iii) indicates that the total irreversibility has two distinct origins: the spontaneous self-organization into a nonequilibrium steady state, and the continuos environmental drive that keeps the system away from its equilibrium.  These terms respectively correspond to the \emph{Boltzmann's thesis} and the \emph{Prigogine's thesis} \cite{Ge:2010fk}. For systems in a non-driving environment, detailed balance holds. Then $Q_{hk}=0$, $\sigma=-dF/dt$, and the system relaxes to an equilibrium steady state with $\sigma=0$. The mathematical theory is an abstraction for an earlier phenomenological study of nonequilibrium steady state thermodynamics by \citet{Oono:1998ly}. 

For almost all applications of stochastic dynamic theories in physics, chemistry and biology, there will be multiple time scales, and often with a significant separation. Recall that in the Gibbs formalism for equilibrium statistical mechanics, the conditional free energy, plays a central role in applications: one usually does not work with the pure mechanical energy of a system; rather, one works with a conditional free energy from coarse-graining and develops a partition function thereafter. The present work focuses on this important issue: \emph{Whether the new thermodynamic structure is invariant in a multiscale stochastic dynamical system}. Or in other words: \emph{whether the relation between state and process variables remains unchanged when the system is viewed at different time scales}. 

We show that the dynamics on a fast time scale contribute an entropic term to the ``internal energy function'', $u_S(x)$, for the slow dynamics. $u_S(x)$ should be understood as the {\em conditional free energy}. And based on $u_S(x)$, one can then treat the dynamics on the slow time scale as if the fast dynamics is nonexistent. Futhermore, we show that the free energy (which characterizes the \emph{spontaneous organization} in a system) is invariant with or without the fast dynamics. Since the dynamics on the fast time scale reaches their stationarity
instantaneously on the slow time scale, they have no effect on the system's free energy. The same can not be said for the entropy and the internal energy, both of which contain the same entropic contribution from the fast dynamics.  Since ``free energy equals internal energy minus entropy'', there is a compensation. Finally, we study how the time-scale separation affects the concepts of stationary and steady-adiabatic processes introduced by \citet{Oono:1998ly}.

\section{Adiabatic approximation}

In this section we follow the ideas of quasi-steady state approximation, or singular perturbation \citep{Qian:2000fk,Rao:2003uq,Pigolotti:2008fk,Zeron:2010fk}, in stochastic dynamics to perform an adiabatic treatment in a generic Markovian stochastic process where separation of time scales is possible. We discover that the concept of conditional probability is a very natural language for performing this multiple time scale analysis. 

Consider a Markov system whose state can be represented by a dual vector $(x,y)$, where variables $x$ and $y$ take discrete values. Let $p(x,y)$ be the probability of state $(x,y)$ and $\upsilon(x,y;x',y')$ be the transition probability per unit time from state $(x,y)$ to state $(x',y')$. We further assume that $\upsilon(x,y;x,y') \gg \upsilon(x'',y'';x''',y''')$ for all $y \neq y'$, $x'' \neq x'''$, and $(x'',y'') \neq (x''',y''')$. 
That is for any given $x$, the transition $y\rightarrow y'$ is much faster than all transitions
involving changing $x''\rightarrow x'''$.
If the sets of all possible values attained by $x$ and $y$ are finite, the master equation (or forward Kolmogorov equation) for this system can be written as
\begin{eqnarray}
  \frac{dp(x,y)}{dt} &=& \sum_{\text{all } y'} p(x,y') \upsilon(x,y';x,y) - p(x,y) \upsilon(x,y;x,y') 
\nonumber\\
  &+& \sum_{x'\neq x}\sum_{\text{all } y'} 
	p(x',y') \upsilon(x',y';x,y) - p(x,y) \upsilon(x,y;x',y').
\label{Meq}
\end{eqnarray}
By adding the above equation over all values of $y$ and taking into consideration that $p(x) = \sum_y p(x,y)$ we obtain
\begin{eqnarray}
 \frac{dp(x)}{dt} &=& \sum_{\text{all } y,y'} p(x,y') \upsilon(x,y';x,y) - p(x,y) \upsilon(x,y;x,y') 
\nonumber\\
  &+& \sum_{x'\neq x} \sum_{\text{all } y,y'} p(x',y') \upsilon(x',y';x,y) - p(x,y) \upsilon(x,y;x',y'). \nonumber
\end{eqnarray}
Note that the first summand in the right hand side of the above equation equals zero because each term in it is added and subtracted once. After some algebra this equation can be rewritten as
\begin{equation}
 \frac{dp(x)}{dt} = \sum_{x'} p(x') \Upsilon(x';x) - p(x) \Upsilon(x;x'),
\label{MeqS}
\end{equation}
with
\begin{equation}
\Upsilon(x;x') = \sum_{y,y'} p(y|x) \upsilon(x,y;x',y').
\label{EffProp}
\end{equation}
In the previous equation $p(y|x)$ is the conditional probability defined as
\begin{equation}
p(y|x) = p(x,y) / p(x).
\label{CondProb}
\end{equation}

In order to find the equation governing the dynamics of $p(y|x)$ let us differentiate (\ref{CondProb}) to obtain the following expression after some algebraic steps: 
\[
p(x) \frac{d p(y|x) }{ dt} = \frac{dp(x,y)}{dt} - p(y|x) \frac{dp(x)}{dt}.
\]
Further substitution of (\ref{Meq}) and (\ref{MeqS}) into this equation leads to
\begin{equation}
 \frac{dp(y|x)}{dt} = \sum_{y'} p(y'|x) \upsilon(x,y';x,y) - p(y|x) \upsilon(x,y;x,y'),
\label{MeqF}
\end{equation}
where we have neglected all terms multiplied by either $\upsilon(x,y;x',y')$ [$x \neq x'$ and $(x,y) \neq (x',y')$] or $\Upsilon(x;x')$, based on the fact that they are much smaller than $\upsilon(x,y';x,y)$.

Finally, from the same assumed time-scale separation, we can make an adiabatic approximation and suppose that $p(y|x) \approx p^s(y|x)$, where the
conditional stationary distribution $p^s(y|x)$ satisfies
\begin{equation}
\sum_{y'} p^s(y'|x) \upsilon(x,y';x,y) - p^s(y|x) \upsilon(x,y;x,y') = 0.
\label{AdAppr}
\end{equation}

In summary, after performing the above described adiabatic approximation, the dynamics of $p(x)$ are governed by (\ref{MeqS}), where the effective transition probability from state $x$ to state $x'$ is given by
\begin{equation}
\Upsilon(x;x') = \sum_{y,y'} p^s(y|x) \upsilon(x,y;x',y'),
\label{EffPropAA}
\end{equation}
while $p^s(y|x)$ is the solution of (\ref{AdAppr}). Notice that the adiabatic approximation that we have introduced in the above paragraphs is equivalent to that introduced by \citet{Pigolotti:2008fk}.

\section{Thermodynamic state functions}

\subsection{Internal energy}

Consider a molecular system that is irreducible; and thus has a unique long-time stationary probability distribution $p^s(x,y)$. Further assume that the system is in contact with an isothermal bath with chemical potential difference. Thus, we can define, following \citet{Ge:2010fk}, the energy function associated to state $(x,y)$ via the stationary distribution $p^s(x,y)$ as
\begin{equation}
u(x,y) = -k_BT \log p^s(x,y),
\label{udef}
\end{equation}
where $k_B$ is the Boltzmann constant and $T$ is the absolute temperature. In systems with detailed balance $p^s(x,y)$ equals the thermodynamic-equilibrium probability distribution $p^e(x,y)$ and Eqn. (\ref{udef}) is equivalent to Gibbs' grand canonical ensemble. When detailed balance is not fulfilled, the above definition of internal energy is related to the \emph{stochastic potential} studied by \citet{Kubo:1973fk}.

From (\ref{udef}), the mean internal energy of the mesoscopic state $p(x,y)$ can be written as
\begin{equation}
U = \sum_{x,y} p(x,y) u(x,y) = -k_BT \sum_{x,y} p(x,y) \log p^s(x,y).
\label{IntEner}
\end{equation}
By substituting (\ref{CondProb}) into (\ref{IntEner}) this last equation can be rearranged as follows:
\begin{equation}
U = \sum_{x,y} p(x,y) u(x,y) = \sum_x p(x) (u_S(x) + u_F(x)),
\label{IntEnerSF}
\end{equation}
where 
\begin{equation}
u_S(x) = -k_BT \log p^s(x) \quad \text{and} \quad u_F(x) = -k_BT \sum_y p(y|x) \log p^s(y|x).
\label{usuf}
\end{equation}
Moreover, if we impose the adiabatic approximation stating that $p(y|x) \approx p^s(y|x)$, 
\begin{equation}
u_F(x) = -k_B T \sum_y p^s(y|x) \log p^s(y|x). 
\label{ufaa}
\end{equation}
These results imply that the internal energy can be split in two components ($U=U_S + U_F$) corresponding to the slow ($U_S = \sum_x p(x) u_S(x)$) and fast ($U_F = \sum_x p(x) u_F(x)$) time scales, respectively.

\subsection{Entropy}

The Gibbs entropy is defined as usual:
\begin{equation}
S = -k_B \sum_{x,y} p(x,y) \log p(x,y).
\label{Entropy}
\end{equation}
Substitution of (\ref{CondProb}) into (\ref{Entropy}) leads to
\begin{equation}
S = -k_B \sum_{x} p(x) \log p(x) -k_B \sum_x p(x) \sum_y p(y|x) \log p(y|x).
\label{EntropySF}
\end{equation}
We see that, once more, the entropy can be separated into slow and fast components $(S=S_S + S_F)$ respectively defined as
\begin{equation}
S_S = -k_B \sum_{x} p(x) \log p(x) \quad \text{and} \quad S_F = -k_B \sum_x p(x) \sum_y p(y|x) \log p(y|x).
\label{sssf}
\end{equation}
If we enforce the adiabatic approximation ($p(y|x)=p^s(y|x)$), the fast component becomes $S_F = \sum_x p(x) s_F(x)$, with
\begin{equation}
s_F(x) = -k_B \sum_y p^s(y|x) \log p^s(y|x) 
\label{sfaa}
\end{equation}
We note by comparing Eqns. (\ref{ufaa}) and (\ref{sfaa}) that $u_F(x)=Ts_F(x)$ due to the adiabatic approximation.

\subsection{Free energy}

From its definition, $F = U-TS$, and Eqns. (\ref{IntEner}) and (\ref{Entropy}), 
the Helmoltz free energy is given by \cite{Ge:2010fk}:
\begin{eqnarray}
	F &=& k_B T \sum_{x,y} p(x,y) \log\left(\frac{p(x,y)}{p^s(x,y)}\right)
\nonumber\\
	&=& k_B T \sum_x p(x) \log \frac{p(x)}{p^s(x)} + k_B T \sum_x p(x) \sum_y p(y|x) \log \frac{p(y|x)}{p^s(y|x)}.
\label{FreeEnerSF}
\end{eqnarray}
In this case it is also possible to identify slow ($F_S$) and fast ($F_F=\sum_x p(x) f_F(x)$) components for the free energy, where
\[
F_S = k_B T \sum_x p(x) \log \frac{p(x)}{p^s(x)} \quad \text{and} \quad f_F(x) = k_B T \sum_y p(y|x) \log \frac{p(y|x)}{p^s(y|x)}.
\]
However, the imposition of the adiabatic approximation implies that $f_F(x)=0$ $\forall x$, and so that $F_F = \sum_x p(x) f_F(x)= 0$. This agrees with the fact that enforcing the adiabatic approximation is equivalent to assuming that the fast time-scale distribution ($p(y|x)$) equilibrates instantaneously with the slow one ($p(x)$) for every given $x$. Therefore, the system's free energy is invariant whether one considers or neglects the faster dynamics, as long as there is a reasonable separation of time scales.

\subsection{Slow-dynamics perspective and whole-system-level interpretation}

First we note from (\ref{ufaa}) and (\ref{sfaa}) that, once the adiabatic approximation has been made, $s_F(x) = u_F(x)/T$. This term should be regarded as the entropy of a state $x$ due to the fast dynamics of variable $y$ within the given $x$. Then, (\ref{IntEnerSF}) indicates that the energy of the slow time scale obeys
\begin{equation}
	u_S(x)= \left(\sum_y p^s(y|x)u(x,y) \right) -Ts_F(x) = \widetilde{u}(x) -Ts_F(x),
\end{equation}
where the first term on the right-hand-side, $\widetilde{u}(x)$, is the mean internal energy of state $x$. Finally, in terms of $\widetilde{u}(x)$, one has the canonical form of the thermodynamics for the slow variable 
\begin{equation}
        F_S = F = \sum_x p(x)\widetilde{u}_S(x)+k_BT\sum_x p(x)\log p(x).
\end{equation}

To gain more insight into the physical meaning of $u_S(x)$ we shall discuss another feasible interpretation for this quantity when $x$ is a continuous variable. In such a case, $u_S(x)$ takes the form of a potential of mean force. In fact, noting that $p^s(x) = \sum_y p^s(x,y)$, together with the definitions for $u(x,y)$ (\ref{udef}) and $u_S(x)$ (\ref{usuf}), one has
\begin{equation}
u_S(x) = - k_B T \log \sum_y \exp ( - u(x,y)/k_B T )
\label{PotMeanF}
\end{equation}
while
\begin{equation}
 \frac{d}{dx} u_S(x) = \frac{\sum_y \exp ( - u(x,y)/k_B T ) \partial u(x,y)/\partial x}
			{\sum_y \exp ( - u(x,y)/k_B T )},
\end{equation}
which corresponds to the usual potential of mean force definition \citep{Kirkwood:1935fk}.

\section{Time evolution and thermodynamic process functions}

\subsection{Time derivative of the thermodynamic functions}

Following \citet{Ge:2010fk}, we shall differentiate the expressions for $U$, $S$, and $F$---Eqns. (\ref{IntEnerSF}), (\ref{EntropySF}), and (\ref{FreeEnerSF})---and write the corresponding rates of change in terms of energy and entropy fluxes; since understanding these fluxes under different conditions provides valuable information regarding the system dynamic and thermodynamic behavior. In particular, we are interested in investigating how the slow and fast dynamics subspaces contribute to the energy and entropy fluxes, and whether their structure remain invariant from the slow-dynamics perspective.

The time derivatives for for $U$, $S$, and $F$ are calculated in Appendix \ref{DotUSF}. After imposing the adiabatic approximation $p(y|x) \approx p^s(y|x)$ on the corresponding expressions we obtain:
\begin{eqnarray}
\dot{U} & = & \displaystyle - \frac{k_B T}{2} \sum_{x,x'} (p(x') \Upsilon(x';x) - p(x) \Upsilon(x;x')) \log \frac{p^s(x)}{p^s(x')}  \nonumber\\
 & & \displaystyle - \frac{k_B T}{2} \sum_x p(x) \sum_{y,y'} (p^s(y'|x) \upsilon(x,y';x,y) - p^s(y|x) \upsilon(x,y;x,y')) \log \frac{p^s(y|x)}{p^s(y'|x)}, \label{dudtAA} \\
\dot{F} & = & - \frac{k_B T}{2} \sum_{x,x'} (p(x') \Upsilon(x';x) - p(x) \Upsilon(x;x')) \log \frac{p(x')p^s(x)}{p(x)p^s(x')}, \label{dfdtAA} \\
\dot{S} & = & \displaystyle \frac{k_B}{2} \sum_{x,x'} (p(x') \Upsilon(x';x) - p(x) \Upsilon(x;x')) \left( \log \frac{p(x') \Upsilon(x';x)}{p(x) \Upsilon(x;x')} - \log \frac{\Upsilon(x';x)}{\Upsilon(x;x')} \right) \nonumber \\
 & & + \frac{k_B }{2} \sum_x p(x) \sum_{y,y'} (p^s(y'|x) \upsilon(x,y';x,y) - p^s(y|x) \upsilon(x,y;x,y')) \nonumber \\
 & & \times \left( \log \frac{p^s(y'|x) \upsilon(x,y';x,y)}{p^s(y|x) \upsilon(x,y;x,y')} -\log \frac{\upsilon(x,y';x,y)}{\upsilon(x,y;x,y')} \right). \label{dsdtAA}
\end{eqnarray}

Before proceeding any further, notice that the formulas for $\dot{U}$ and $\dot{S}$ posses terms corresponding to the slow and fast dynamics subspaces. Moreover, the slow and fast dynamics terms in each equation have the same general structure. The same is true when each (slow or fast dynamics) term is compared with that on the right hand side of the corresponding equation in \citep{Ge:2010fk}. Finally, because of the adiabatic approximation, the fast-dynamics  terms in $\dot{U}$ and $\dot{S}$ are equal, except for the multiplicative factor $T$. Hence, they cancel in $U - TS$ and, in consequence, the time derivative for the free energy ($\dot{F}$) is the same no matter wether a fast time scale exists or not \cite{Qian:2001fk}. 

\subsection{Detailed balance}

So far, we have obtained all of our results without making use of the detailed balance condition. When the environment of a stochastic system is not driving it out of equilibrium, the system ultimately reaches an equilibrium steady state which is characterized by the fulfillment of detailed balance:
\begin{equation}
	p^e(x,y)\upsilon(x,y;x',y') = p^e(x',y')\upsilon(x',y';x,y).
\label{DetBal}
\end{equation}
Through the present section we denote the stationary distribution as $p^e(x,y)$, rather than $p^s(x,y)$, to emphasize the fact that it obeys detailed balance and thus corresponds to thermodynamic equilibrium.

Consider the effective transition probability defined in (\ref{EffProp}) and make use 
of (\ref{CondProb}) to arrive at the following expression:
\[
p(x)\Upsilon(x;x') = \sum_{y,y'} p(x,y) \upsilon(x,y;x',y').
\]
Assume now that the system is in equilibrium and substitute Eqn. (\ref{DetBal}) into the above equation to obtain
\begin{equation}
	p^e(x)\Upsilon(x;x') = p^e(x')\Upsilon(x';x).
\label{DetBalAA}
\end{equation}
That is, Eq. (\ref{DetBalAA}) is the form of the detailed balance condition for
the variable with slow dynamics, with the probability distribution $p^e(x) = \sum_y p^e(x,y)$. On the other hand for the fast dynamic variable, it follows from (\ref{CondProb}) that detailed balance implies that
\begin{equation}
	p^e(y|x)\upsilon(x,y;x,y') = p^e(y'|x)\upsilon(x,y';x,y).
\label{DetBalFD}
\end{equation}

By employing the above results and following the procedure introduced by \citet{Ge:2010fk}, we can decompose $\dot{U}$, $\dot{S}$, and $\dot{F}$ as follows:
\begin{equation}
\dot{U} = - Q_d, \quad \dot{F} = - T \sigma, \quad \text{and} \quad \dot{S} = \sigma - \frac{Q_d}{T},
\label{SUFdot}
\end{equation}
with
\begin{eqnarray}
Q_d & = & - \dot{U} = \frac{k_B T}{2} \sum_{x,x'} (p(x') \Upsilon(x';x) - p(x) \Upsilon(x;x')) \log \frac{\Upsilon(x';x)}{\Upsilon(x;x')}, \label{HeatDiss} \\
\sigma & = & \frac{k_B}{2} \sum_{x,x'} (p(x') \Upsilon(x';x) - p(x) \Upsilon(x;x')) \log \frac{p(x') \Upsilon(x';x)}{p(x) \Upsilon(x;x')}. \label{EntProd}
\end{eqnarray}
A comparison of Eqns. (\ref{SUFdot}) and (\ref{EntProd}) with the equations defining the dissipation heat and the entropy production rate in \citep{Ge:2010fk}, respectively, reveals that $Q_d$ and $\sigma$ posses the same mathematical structure as, and thus can be identified with those quantities.

When the system is in equilibrium with detailed balance, $\dot{U} = \dot{F} = \dot{S} = 0$. Furthermore, it is straightforward to verify that $Q_d = \sigma = 0$ as well. We thus conclude from these results that the thermodynamic equilibrium state is characterized not only by the constancy in time of the thermodynamic state functions $U$, $F$, and $S$, but also by the existence of neither an energy flow nor an entropy production.

\subsection{Process functions for systems without detailed balance}

We are now going back to Eqns. (\ref{dudtAA})-(\ref{dsdtAA}). We see by following the procedure in \citep{Ge:2010fk} that, when detailed balance is not fulfilled, the entropy rate of change can still be decomposed as
\begin{equation}
\dot{S} = \sigma - \frac{Q_d}{T},
\label{SdotNDB}
\end{equation}
where the entropy production rate is now given by
\begin{eqnarray}
\sigma & = & \frac{k_B}{2} \sum_{x,x'} (p(x') \Upsilon(x';x) - p(x) \Upsilon(x;x')) \log \frac{p(x') \Upsilon(x';x)}{p(x) \Upsilon(x;x')} \nonumber \\
 & & + \frac{k_B}{2} \sum_x p(x) \sum_{y,y'} (p^s(y'|x) \upsilon(x,y';x,y) - p^s(y|x) \upsilon(x,y;x,y')) \nonumber \\
 & & \times \log \frac{p^s(y'|x)\upsilon(x,y';x,y)}{p^s(y|x)\upsilon(x,y;x,y')}. \nonumber
\end{eqnarray}
while the dissipated heat rate is
\begin{eqnarray}
Q_d & = & \frac{k_B T}{2} \sum_{x,x'} (p(x') \Upsilon(x';x) - p(x) \Upsilon(x;x')) \log \frac{\Upsilon(x';x)}{\Upsilon(x;x')} \nonumber \\
 & & + \frac{k_B T}{2} \sum_x p(x) \sum_{y,y'} (p^s(y'|x) \upsilon(x,y';x,y) - p^s(y|x) \upsilon(x,y;x,y')) \nonumber \\
 & & \times \log \frac{\upsilon(x,y';x,y)}{\upsilon(x,y;x,y')}.
\label{HeatDissNDB}
\end{eqnarray}
Observe that both $\sigma$ and $Q_d$ can be decomposed into two different terms with the same structure, each one of them corresponding to the slow and fast dynamics subspaces.

Eqn. (\ref{SdotNDB}) is one of the fundamental postulates of phenomenological 
irreversible thermodynamics \cite{deGroot:1951}.
Using these definitions we can also rewrite $\dot{U}$ and $\dot{F}$ as
\begin{equation}
\dot{U} = Q_{hk} - Q_d \quad \text{and} \quad \dot{F} = Q_{hk} - T \sigma ,
\label{UFdotNDB}
\end{equation}
where
\begin{eqnarray}
Q_{hk} & = & \frac{k_B T}{2} \sum_{x,x'} (p(x') \Upsilon(x';x) - p(x) \Upsilon(x;x')) \log \frac{p^s(x') \Upsilon(x';x)}{p^s(x) \Upsilon(x;x')} \nonumber \\
 & & + \frac{k_B T}{2} \sum_x p(x) \sum_{y,y'} (p^s(y'|x) \upsilon(x,y';x,y) - p^s(y|x) \upsilon(x,y;x,y')) \nonumber \\
 & & \times \log \frac{p^s(y'|x)\upsilon(x,y';x,y)}{p^s(y|x)\upsilon(x,y;x,y')}. \nonumber
\end{eqnarray}
This expression for $Q_{hk}$ can again be decomposed into two terms corresponding to the slow and fast dynamics subspaces. Observe that both terms have the same mathematical structure as the definition for the housekeeping heat in \citep{Ge:2010fk}. Hence, we can identify $Q_{hk}$ with this quantity, originally introduced by Oono and Paniconi \cite{Oono:1998ly,Ge:2010fk} and interpreted as the energy flow that has to be administered to the system to keep the stationary state out of equilibrium. 

Define now 
\begin{eqnarray}
A(x,y',y) & = & (p^s(y'|x) \upsilon(x,y';x,y) - p^s(y|x) \upsilon(x,y;x,y')), \nonumber \\
B(x,y) & = & \log p^s(y|x), \nonumber 
\end{eqnarray} 
It is straightforward to verify that $A$ is antisymmetric in $y$ and $y'$: $A(x,y',y) = - A(x,y,y')$. Moreover, since $p^s(y|x)$ is by definition the stationary conditional probability distribution for variable $y$ (conditioned to the value of $x$), it follows from Eqn. (\ref{MeqF}) that $\sum_{y'} A(x,y',y) = 0$ $\forall x,y$.  Furthermore, as a function
of $y$ and $y'$, $A$ is an antisymmetric matrix with all its rows, thus all columns,
summing zero.  Then for any real vector $B$ with component $B(\cdot,y)$:
\begin{eqnarray}
	&& \sum_{y,y'} A(x,y,y') \left(B(x,y)-B(x,y')\right) 
\nonumber\\
	&=&
	\sum_{y} B(x,y)\left( \sum_{y'} A(x,y,y')\right)  
		- \sum_{y'}B(x,y') \left(\sum_{y} A(x,y,y')\right) = 0.
\label{lemma1}
\end{eqnarray}
This result further implies that 
\begin{eqnarray}
Q_{hk} & = & \frac{k_B T}{2} \sum_{x,x'} (p(x') \Upsilon(x';x) - p(x) \Upsilon(x;x')) \log \frac{p^s(x') \Upsilon(x';x)}{p^s(x) \Upsilon(x;x')} \
\nonumber\\
 & & + \frac{k_B T}{2} \sum_x p(x) \sum_{y,y'} (p^s(y'|x) \upsilon(x,y';x,y) - p^s(y|x) \upsilon(x,y;x,y')) \nonumber \\
 & & \times \log \frac{\upsilon(x,y';x,y)}{\upsilon(x,y;x,y')}, 
\label{HeatHK} \\
\sigma & = & \frac{k_B}{2} \sum_{x,x'} (p(x') \Upsilon(x';x) - p(x) \Upsilon(x;x')) \log \frac{p(x') \Upsilon(x';x)}{p(x) \Upsilon(x;x')} \nonumber \\
 & & + \frac{k_B}{2} \sum_x p(x) \sum_{y,y'} (p^s(y'|x) \upsilon(x,y';x,y) - p^s(y|x) \upsilon(x,y;x,y')). \nonumber \\
 & & \times \log \frac{\upsilon(x,y';x,y)}{\upsilon(x,y;x,y')}. \label{EntProdNDB}
\end{eqnarray}
Finally, the expression for $\dot{U}$ and $\dot{S}$ transform into
\begin{eqnarray}
\dot{U} & = & \displaystyle - \frac{k_B T}{2} \sum_{x,x'} (p(x') \Upsilon(x';x) - p(x) \Upsilon(x;x')) \log \frac{p^s(x)}{p^s(x')}, \label{dudtNDB} \\
\dot{S} & = & \displaystyle \frac{k_B}{2} \sum_{x,x'} (p(x') \Upsilon(x';x) - p(x) \Upsilon(x;x')) \nonumber \\
& & \times \left( \log \frac{p(x') \Upsilon(x';x)}{p(x) \Upsilon(x;x')} - \log \frac{\Upsilon(x';x)}{\Upsilon(x;x')} \right), \label{dsdtNDB}
\end{eqnarray}
while $\dot{F}$ remains the same as in Eqn. (\ref{dfdtAA}).

Let us define
\begin{equation}
Q_{fast} = \frac{k_B T}{2} \sum_x p(x) \sum_{y,y'} (p^s(y'|x) \upsilon(x,y';x,y) - p^s(y|x) \upsilon(x,y;x,y')) \log \frac{\upsilon(x,y';x,y)}{\upsilon(x,y;x,y')}.
\label{Qfast}
\end{equation}
We can see from this definition that $Q_{fast}$ is an energy flux related to fast time scale. Observe that $Q_{fast}$ appears as a summand in the expressions for $Q_d$ (\ref{HeatDissNDB}) and $Q_{hk}$ (\ref{HeatHK}), while $Q_{fast}/T$ appears in the expression for $\sigma$ (\ref{EntProdNDB}). That is, the fast dynamics contributions to the dissipated heat, the housekeeping heat, and the entropy production rate are identical (except for a factor $T$ in the case of $\sigma$) in all three cases. Furthermore, $Q_{fast}$ cancels when $Q_d$, $Q_{hk}$, and $T\sigma$ are subtracted and this explains why such term does not appear in the expressions for $\dot{U}$, $\dot{S}$, and $\dot{F}$.

\subsection{Partial detailed balance with rapid pre-equilibrium}

Assume that detailed balance is fulfilled by the fast dynamics distribution ($p(y|x)$), but not necessarily by $p(x)$. Then there is a rapid pre-equilibrium 
$p(y|x) \approx p^e(y|x)$, with $p^e(y|x)$ satisfying Eqn. (\ref{DetBalFD}). If this is the case, then $Q_{fast} = 0$. Interestingly, the expressions for $\dot{U}$, $\dot{S}$, and $\dot{F}$ do not change. They are the same as in Eqns. (\ref{dfdtAA}), (\ref{dudtNDB}), and (\ref{dsdtNDB}), except that $p^s(y|x)$ is substituted by $p^e(y|x)$ whenever the former term appears. All this means that, having or not having detailed balance in the fast dynamics space makes a difference for the energy flows $Q_d$ and $Q_{hk}$, as well as for the entropy production rate $\sigma$ (all of them are smaller in the first case because the contribution due to fast dynamics vanishes), however this difference is transparent to the rate of change of all thermodynamic state functions ($U$, $S$, and $F$).

\subsection{Stationary distribution without detailed balance}

To analyze the behavior of the process variables when the system is at a steady state without detailed balance let us define
\begin{eqnarray}
\mathcal{A}(x,x') & = & p^s(x) \Upsilon(x;x') - p^s('x) \Upsilon(x';x), \nonumber \\
\mathcal{B}(x) & = & \log p^s(x), \nonumber \\
\mathcal{C}(x) & = & u^s(x) / k_B T. \nonumber
\end{eqnarray}
Clearly, $\mathcal{A}(x,x')$ is antisymmetric ($\mathcal{A}(x,x') = - \mathcal{A}(x',x)$). Moreover, since $p^s(x)$ is the stationary probability distribution for variable $x$, it follows from Eqn. (\ref{MeqS}) that $\sum_{x'} A(x',x) = 0$. Hence, by the same
result in Eq. (\ref{lemma1}) we have
\[
\sum_{x,x'} \mathcal{A}(x',x) (\mathcal{B}(x')-\mathcal{B}(x)) = \sum_{x,x'} \mathcal{A}(x',x) (\mathcal{C}(x')-\mathcal{C}(x)) = 0.
\]
This last equation, together with (\ref{dfdtAA}), (\ref{dudtNDB}), and (\ref{dsdtNDB}) further imply that 
\begin{equation}
T \sigma = Q_d = Q_{hk} = \frac{k_B T}{2} \sum_{x,x'} (p^s(x') \Upsilon(x';x) - p^s(x) \Upsilon(x;x')) \log \frac{\Upsilon(x';x)}{\Upsilon(x;x')} + Q_{fast},
\label{StatFlux}
\end{equation}
when $p(x) = p^s(x)$ and $p(y|x) = p^s(y|x)$. Finally, it results from Eqn. (\ref{StatFlux}) that $\dot{U} = \dot{S} =\dot{F} = 0$ in the stationary state. Indeed, we can see from (\ref{FreeEnerSF}) that $F = 0$ in such case. 

The results in the foregoing paragraph corroborate the following: once the system reaches the steady state distribution, all the thermodynamic state functions (internal energy, free energy, and entropy) will remain constant. However, contrary to an equilibrium steady state in which detailed balance is fulfilled, a nonequilibrium steady state has nonzero fluxes given by (\ref{StatFlux}). The equalities between the fluxes reflect both the energy conservation and the isothermal Clausius equality: on the one hand, to keep the system out-of-equilibrium, energy has to be supplied to the system ($Q_{hk}$) which is then dissipated as heat ($Q_d$); while, on the other hand, entropy is produced in the process of the conversion of useful energy to heat ($\sigma=Q_d/T$).

\section{Quasi-stationary and steady-adiabatic processes}

The concepts of quasi-stationary and adiabatic processes are central to thermodynamics. In systems where the stationary state satisfies detailed balance, a quasi-stationary process can be defined as a succession of states where $\sigma=0$, while an adiabatic processes is a succession of states satisfying $Q_d = 0$.

\citet{Oono:1998ly} generalized these concepts for systems with a non-equilibrium steady state (NESS) by defining the excess heat and the free energy dissipation rate as follows:
\begin{eqnarray}
Q_{ex} & = & Q_d - Q_{hk}, \label{qexdef} \\ 
\theta & = & T\sigma - Q_{hk}, \label{thetadef}
\end{eqnarray}
and noting that, in terms of these variables, the rates $\dot{S}$, $\dot{U}$, and $\dot{F}$ can be rewritten as---see Eqns. (\ref{SdotNDB}) and (\ref{UFdotNDB}):
\begin{equation}
\dot{S} = \frac{\theta - Q_{ex}}{T}, \quad \dot{U} = -Q_{ex}, \quad \dot{F} = - \theta.
\label{SUFdotNDB}
\end{equation}
A comparison of Eqns. (\ref{SUFdot}) and (\ref{SUFdotNDB}) reveals that $Q_d$ and $\sigma$ in systems where the stationary state satisfies detailed balance can be respectively identified with $Q_{ex}$ and $\theta/T$ in NESS systems. Based on this identification \citet{Oono:1998ly} generalized the concepts of quasi-stationary and steady-adiabatic processes for NESS systems as follows: a quasi-stationary process is a succession of states satisfying $\theta = 0$, while a steady-adiabatic process is a succession of states complying with $Q_{ex} = 0$. After introducing these concepts, \citet{Oono:1998ly} made extensive use of them in the development of their phenomenological steady-state thermodynamics. Here we investigate how a time-scale separation affects these processes.

After substituting (\ref{HeatDissNDB}), (\ref{HeatHK}) and (\ref{EntProdNDB}) into (\ref{qexdef}) and (\ref{thetadef})  we obtain
\begin{eqnarray}
Q_{ex} & = & \frac{k_B T}{2} \sum_{x,x'} (p(x') \Upsilon(x';x) - p(x) \Upsilon(x;x')) \log \frac{p^s(x)}{p^s(x')}, \label{qex} \\
\theta & = & \frac{k_B}{2} \sum_{x,x'} (p(x') \Upsilon(x';x) - p(x) \Upsilon(x;x')) \log \frac{p(x') p^s(x)}{p(x) p^s(x')}. \label{theta}
\end{eqnarray}
Recall that the energy flux related to the fast time scale $Q_{fast}$---see Eqn. (\ref{Qfast})---appears as a summand in $Q_d$, $Q_{hk}$, and $\sigma$. Hence, it cancels out at the time of subtracting these quantities and so it shows neither in $Q_{ex}$ nor in $\theta$. Consequently, $\theta$ has no contribution whatsoever from the fast dynamics subspace.

The fact that $\theta$ depends only on the slow-dynamics subspace $x$ means that the fast-dynamics subspace ($y$) does not influence whether a given process is quasi-stationary or not. This result is in agreement with the adiabatic approximation we have made to reduce the system master equation, which is equivalent to assuming that the fast-dynamics subspace immediately equilibrates with the slow-dynamics state $x$.

Regarding steady-adiabatic processes for NESS systems we see that, since $Q_{ex}$ depends on the fast dynamics through $s_F(x)$---see Eqns. (\ref{sfaa}) and (\ref{qex}), the fast dynamics cannot be ignored while determining the adiabaticity of a given process. This can be more clearly appreciated  by noticing that the following generalized Clausius equality is satisfied in a quasi-stationary process:
\[
\dot{S} = - \frac{Q_{ex}}{T}.
\]

\section{Concluding remarks}

In this work we have extended the results of a recent paper \citep{Ge:2010fk} where a rather complete thermodynamic formalism was introduced for discrete-state, continuous-time Markov processes with and without detailed balance. Our main objective was to investigate whether the thermodynamic structure is invariant in a multiscale stochastic system. By invariance we mean that the relation between state and process variables remains unchanged when the system is viewed at different time scales.

We proceeded as follows. First, we assumed that the states of a system can be classified according to the propensities of the transitions among them. More precisely, we supposed that every state can be represented by a dual vector $(x,y)$, and that transitions involving changes in $y$ alone are much more probable than those involving changes in $x$ or in both $x$ and $y$. Then, we imposed an adiabatic approximation to deduce a reduced master equation for the slower time scale. Finally, we analyzed the implications of this adiabatic approximation on the thermodynamic formalism introduced by \citet{Ge:2010fk}.

As it resulted, all thermodynamic variables and their time derivatives can be separated in a very natural way into contributions from the slow and fast time scales. The only exceptions being the Helmholtz free energy and its time derivatives, which only involve terms due to the slower time scale. In other words, the Helmholtz free energy (which characterizes the system spontaneous organization) is invariant with and without a fast time scale. This happens because, having reached its stationarity, the fast time-scale probability distribution does not contribute to the free energy. The same cannot be said about the entropy and the internal energy, both of which have fast dynamics contributions. 

The above discussed results are important because they provide a framework to study the thermodynamics of complex Markov processes where time-scale separation is possible. Some examples where this framework can be useful may be enzymatic reactions in which one of the chemical steps is much slower than the rest, of gene regulatory networks where, typically, transcription initiation is an infrequent process, as compared for instance with translation initiation and post-translational modification of the resulting proteins.

On the other hand, the same results can also be interpreted from a more fundamental thermodynamic perspective. We elaborate on these ideas next. The dynamics on a fast time scale contribute an \emph{entropic} term to the energy function $u_S(x)$, for the slow dynamics. This fact allows to identify $u_S(x)$ as a \emph{free energy}. Indeed, $u_S(x)$ results to be the conditional free energy; a concept extensively used in equilibrium thermodynamics as one does not usually work with pure mechanical energy, but with a coarse grained conditional free energy, and develop a partition function thereafter.

On the other hand, entropy-enthalpy compensation has been extensively studied in biochemistry \citep{Qian:1996uq, Qian:1998kx}. The strong form of this phenomenon occurs when variations in $\Delta H$ and $\Delta S$, caused by regular changes in some experimental variable (excluding temperature), exhibit a linear correlation. In this case $\Delta G$ will be small relative to the range of values expected from the experiment.

As pointed out by \citet{Qian:2001fk}, internal energy is the equivalent of enthalpy, while Helmholtz free energy is the equivalent of Gibbs free energy, in the type of systems here studied. In that respect, entropy-internal energy compensation in these systems is tantamount to entropy-enthalpy compensation in isobaric ones; and the effect of entropy-internal energy compensation will be small $F$ changes. The insight from the present work is that the compensating part of entropy and internal energy is the contribution from fast dynamics; i.e., rapid fluctuations.

The expressions we derived for $dU/dt$ and $dS/dt$ contain terms associated to the slow and fast time scales. However, when the adiabatic approximation is imposed, the contributions from the faster time scale become equal, except for a factor $T$ in $dS/dt$, see Eqns. (\ref{dudtAA}) and (\ref{dsdtAA}). As a consequence, the expression for $dF/dt$ only includes a slow dynamics term---see Eqn. (\ref{dfdtAA}). That is, we have entropy-internal energy cancelation for the fast dynamics contributions. As a matter of fact, it is impossible to know whether a fast time scale exists or not from the perspective of the Helmholtz free energy. These results are in complete agreement with previous studies which prove the existence of entropy-enthalpy compensation by considering that, in response to a small perturbation, the free energy change of an stationary system is independent of the system thermodynamic environment, while the entropy and the internal energy changes depend on the environmental constraints \citep{Qian:1996uq,Qian:1998kx}. Recall that the adiabatic approximation is equivalent to assuming that the fast dynamics distribution $p(y|x)$ reaches its stationary value instantaneously for every state $x(t)$

\acknowledgements

This work was partially supported by Conacyt, M\'{e}xico, Grant: 55228. The authors are thankful to the anonymous referees, whose comments and suggestions greatly help us to improve the paper.

\bibliography{MSIrrTherm}

\begin{thebibliography}{27}
\expandafter\ifx\csname natexlab\endcsname\relax\def\natexlab#1{#1}\fi
\expandafter\ifx\csname bibnamefont\endcsname\relax
  \def\bibnamefont#1{#1}\fi
\expandafter\ifx\csname bibfnamefont\endcsname\relax
  \def\bibfnamefont#1{#1}\fi
\expandafter\ifx\csname citenamefont\endcsname\relax
  \def\citenamefont#1{#1}\fi
\expandafter\ifx\csname url\endcsname\relax
  \def\url#1{\texttt{#1}}\fi
\expandafter\ifx\csname urlprefix\endcsname\relax\def\urlprefix{URL }\fi
\providecommand{\bibinfo}[2]{#2}
\providecommand{\eprint}[2][]{\url{#2}}

\bibitem[{\citenamefont{Evans et~al.}(1993)\citenamefont{Evans, Cohen, and
  Morriss}}]{Evans:1993fk}
\bibinfo{author}{\bibfnamefont{D.~J.} \bibnamefont{Evans}},
  \bibinfo{author}{\bibfnamefont{E.~G.~D.} \bibnamefont{Cohen}},
  \bibnamefont{and} \bibinfo{author}{\bibfnamefont{G.~P.}
  \bibnamefont{Morriss}}, \bibinfo{journal}{Phys. Rev. Lett.}
  \textbf{\bibinfo{volume}{71}}, \bibinfo{pages}{2401} (\bibinfo{year}{1993}).

\bibitem[{\citenamefont{Evans and Searles}(1994)}]{Evans:1994uq}
\bibinfo{author}{\bibfnamefont{D.~J.} \bibnamefont{Evans}} \bibnamefont{and}
  \bibinfo{author}{\bibfnamefont{D.~J.} \bibnamefont{Searles}},
  \bibinfo{journal}{Phys. Rev. E} \textbf{\bibinfo{volume}{50}},
  \bibinfo{pages}{1645} (\bibinfo{year}{1994}).

\bibitem[{\citenamefont{Gallavotti and Cohen}(1995)}]{Gallavotti:1995}
\bibinfo{author}{\bibfnamefont{G.}~\bibnamefont{Gallavotti}} \bibnamefont{and}
  \bibinfo{author}{\bibfnamefont{E.~G.~D.} \bibnamefont{Cohen}},
  \bibinfo{journal}{Phys. Rev. Lett.} \textbf{\bibinfo{volume}{74}},
  \bibinfo{pages}{2694} (\bibinfo{year}{1995}).

\bibitem[{\citenamefont{Kurchan}(1998)}]{kurchan:1998}
\bibinfo{author}{\bibfnamefont{J.}~\bibnamefont{Kurchan}}, \bibinfo{journal}{J.
  Phys. A: Math. Gen.} \textbf{\bibinfo{volume}{31}}, \bibinfo{pages}{3719}
  (\bibinfo{year}{1998}).

\bibitem[{\citenamefont{Lebowitz and Spohn}(1999)}]{lebowitz}
\bibinfo{author}{\bibfnamefont{J.~L.} \bibnamefont{Lebowitz}} \bibnamefont{and}
  \bibinfo{author}{\bibfnamefont{H.}~\bibnamefont{Spohn}}, \bibinfo{journal}{J.
  Stat. Phys.} \textbf{\bibinfo{volume}{95}}, \bibinfo{pages}{333}
  (\bibinfo{year}{1999}).

\bibitem[{\citenamefont{Crooks}(1999)}]{crooks}
\bibinfo{author}{\bibfnamefont{G.~E.} \bibnamefont{Crooks}},
  \bibinfo{journal}{Phys. Rev. E} \textbf{\bibinfo{volume}{60}},
  \bibinfo{pages}{2721} (\bibinfo{year}{1999}).

\bibitem[{\citenamefont{Wang et~al.}(2002)\citenamefont{Wang, Sevick, Mittag,
  Searles, and Evans}}]{Wang:2002kx}
\bibinfo{author}{\bibfnamefont{G.~M.} \bibnamefont{Wang}},
  \bibinfo{author}{\bibfnamefont{E.~M.} \bibnamefont{Sevick}},
  \bibinfo{author}{\bibfnamefont{E.}~\bibnamefont{Mittag}},
  \bibinfo{author}{\bibfnamefont{D.~J.} \bibnamefont{Searles}},
  \bibnamefont{and} \bibinfo{author}{\bibfnamefont{D.~J.} \bibnamefont{Evans}},
  \bibinfo{journal}{Phys. Rev. Lett.} \textbf{\bibinfo{volume}{89}},
  \bibinfo{pages}{050601} (\bibinfo{year}{2002}).

\bibitem[{\citenamefont{Qian et~al.}(2002)\citenamefont{Qian, Qian, and
  Tang}}]{Qian:2002ys}
\bibinfo{author}{\bibfnamefont{H.}~\bibnamefont{Qian}},
  \bibinfo{author}{\bibfnamefont{M.}~\bibnamefont{Qian}}, \bibnamefont{and}
  \bibinfo{author}{\bibfnamefont{X.}~\bibnamefont{Tang}},
  \bibinfo{journal}{Journal of Statistical Physics}
  \textbf{\bibinfo{volume}{107}}, \bibinfo{pages}{1129} (\bibinfo{year}{2002}).

\bibitem[{\citenamefont{Ge}(2009)}]{Ge:2009}
\bibinfo{author}{\bibfnamefont{H.}~\bibnamefont{Ge}}, \bibinfo{journal}{Phys.
  Rev. E} \textbf{\bibinfo{volume}{80}}, \bibinfo{pages}{021137}
  (\bibinfo{year}{2009}).

\bibitem[{\citenamefont{Ge and Qian}(2010)}]{Ge:2010fk}
\bibinfo{author}{\bibfnamefont{H.}~\bibnamefont{Ge}} \bibnamefont{and}
  \bibinfo{author}{\bibfnamefont{H.}~\bibnamefont{Qian}},
  \bibinfo{journal}{Phys. Rev. E} \textbf{\bibinfo{volume}{81}},
  \bibinfo{pages}{051133} (\bibinfo{year}{2010}).

\bibitem[{\citenamefont{Esposito and Van~den
  Broeck}(2010{\natexlab{a}})}]{esposito_prl:2010}
\bibinfo{author}{\bibfnamefont{M.}~\bibnamefont{Esposito}} \bibnamefont{and}
  \bibinfo{author}{\bibfnamefont{C.}~\bibnamefont{Van~den Broeck}},
  \bibinfo{journal}{Phys. Rev. Lett.} \textbf{\bibinfo{volume}{104}},
  \bibinfo{pages}{090601} (\bibinfo{year}{2010}{\natexlab{a}}).

\bibitem[{\citenamefont{Esposito and Van~den
  Broeck}(2010{\natexlab{b}})}]{esposito_pre1:2010}
\bibinfo{author}{\bibfnamefont{M.}~\bibnamefont{Esposito}} \bibnamefont{and}
  \bibinfo{author}{\bibfnamefont{C.}~\bibnamefont{Van~den Broeck}},
  \bibinfo{journal}{Phys. Rev. E} \textbf{\bibinfo{volume}{82}},
  \bibinfo{pages}{011143} (\bibinfo{year}{2010}{\natexlab{b}}).

\bibitem[{\citenamefont{Van~den Broeck and
  Esposito}(2010)}]{esposito_pre2:2010}
\bibinfo{author}{\bibfnamefont{C.}~\bibnamefont{Van~den Broeck}}
  \bibnamefont{and} \bibinfo{author}{\bibfnamefont{M.}~\bibnamefont{Esposito}},
  \bibinfo{journal}{Phys. Rev. E} \textbf{\bibinfo{volume}{82}},
  \bibinfo{pages}{011144} (\bibinfo{year}{2010}).

\bibitem[{\citenamefont{Bergmann and Lebowitz}(1955)}]{Lebowitz:1955}
\bibinfo{author}{\bibfnamefont{P.}~\bibnamefont{Bergmann}} \bibnamefont{and}
  \bibinfo{author}{\bibfnamefont{J.}~\bibnamefont{Lebowitz}},
  \bibinfo{journal}{Phys. Rev.} \textbf{\bibinfo{volume}{99}},
  \bibinfo{pages}{578} (\bibinfo{year}{1955}).

\bibitem[{\citenamefont{Qian}(2001{\natexlab{a}})}]{Qian:2001}
\bibinfo{author}{\bibfnamefont{H.}~\bibnamefont{Qian}}, \bibinfo{journal}{Phys.
  Rev. E} \textbf{\bibinfo{volume}{63}}, \bibinfo{pages}{042103}
  (\bibinfo{year}{2001}{\natexlab{a}}).

\bibitem[{\citenamefont{Mackey}(2003)}]{Mackey:2003zr}
\bibinfo{author}{\bibfnamefont{M.~C.} \bibnamefont{Mackey}},
  \emph{\bibinfo{title}{Time's {A}rrow: {T}he {O}rigins of {T}hermodynamic
  {B}ehaviour}} (\bibinfo{publisher}{Dover Publications},
  \bibinfo{address}{Mineola, N.Y.}, \bibinfo{year}{2003}).

\bibitem[{\citenamefont{Oono and Paniconi}(1998)}]{Oono:1998ly}
\bibinfo{author}{\bibfnamefont{Y.}~\bibnamefont{Oono}} \bibnamefont{and}
  \bibinfo{author}{\bibfnamefont{M.}~\bibnamefont{Paniconi}},
  \bibinfo{journal}{Prog. Theor. Phys. Suppl.} \textbf{\bibinfo{volume}{130}},
  \bibinfo{pages}{29} (\bibinfo{year}{1998}).

\bibitem[{\citenamefont{Qian}(2000)}]{Qian:2000fk}
\bibinfo{author}{\bibfnamefont{H.}~\bibnamefont{Qian}},
  \bibinfo{journal}{Journal of Mathematical Chemistry}
  \textbf{\bibinfo{volume}{27}}, \bibinfo{pages}{219} (\bibinfo{year}{2000}).

\bibitem[{\citenamefont{Rao and Arkin}(2003)}]{Rao:2003uq}
\bibinfo{author}{\bibfnamefont{C.}~\bibnamefont{Rao}} \bibnamefont{and}
  \bibinfo{author}{\bibfnamefont{A.}~\bibnamefont{Arkin}},
  \bibinfo{journal}{Journal of Chemical Physics}
  \textbf{\bibinfo{volume}{118}}, \bibinfo{pages}{4999} (\bibinfo{year}{2003}).

\bibitem[{\citenamefont{Pigolotti and Vulpiani}(2008)}]{Pigolotti:2008fk}
\bibinfo{author}{\bibfnamefont{S.}~\bibnamefont{Pigolotti}} \bibnamefont{and}
  \bibinfo{author}{\bibfnamefont{A.}~\bibnamefont{Vulpiani}},
  \bibinfo{journal}{The Journal of Chemical Physics}
  \textbf{\bibinfo{volume}{128}}, \bibinfo{eid}{154114}
  (pages~\bibinfo{numpages}{8}) (\bibinfo{year}{2008}).

\bibitem[{\citenamefont{Zeron and Santill\'an}(2010)}]{Zeron:2010fk}
\bibinfo{author}{\bibfnamefont{E.~S.} \bibnamefont{Zeron}} \bibnamefont{and}
  \bibinfo{author}{\bibfnamefont{M.}~\bibnamefont{Santill\'an}},
  \bibinfo{journal}{Journal of Theoretical Biology} \textbf{\bibinfo{volume}{In
  Press}} (\bibinfo{year}{2010}).

\bibitem[{\citenamefont{Kubo et~al.}(1973)\citenamefont{Kubo, Matsuo, and
  Kitahara}}]{Kubo:1973fk}
\bibinfo{author}{\bibfnamefont{R.}~\bibnamefont{Kubo}},
  \bibinfo{author}{\bibfnamefont{K.}~\bibnamefont{Matsuo}}, \bibnamefont{and}
  \bibinfo{author}{\bibfnamefont{K.}~\bibnamefont{Kitahara}},
  \bibinfo{journal}{Journal of Statistical Physics}
  \textbf{\bibinfo{volume}{9}}, \bibinfo{pages}{51} (\bibinfo{year}{1973}),
  ISSN \bibinfo{issn}{0022-4715}, \bibinfo{note}{10.1007/BF01016797},
  \urlprefix\url{http://dx.doi.org/10.1007/BF01016797}.

\bibitem[{\citenamefont{Kirkwood}(1935)}]{Kirkwood:1935fk}
\bibinfo{author}{\bibfnamefont{J.~G.} \bibnamefont{Kirkwood}},
  \bibinfo{journal}{Journal of Chemical Physics} \textbf{\bibinfo{volume}{3}},
  \bibinfo{pages}{300} (\bibinfo{year}{1935}).

\bibitem[{\citenamefont{Qian}(2001{\natexlab{b}})}]{Qian:2001fk}
\bibinfo{author}{\bibfnamefont{H.}~\bibnamefont{Qian}}, \bibinfo{journal}{Phys.
  Rev. E} \textbf{\bibinfo{volume}{65}}, \bibinfo{pages}{016102}
  (\bibinfo{year}{2001}{\natexlab{b}}).

\bibitem[{\citenamefont{de~Groot}(1951)}]{deGroot:1951}
\bibinfo{author}{\bibfnamefont{S.~R.} \bibnamefont{de~Groot}},
  \emph{\bibinfo{title}{Thermodynamics of {I}rreversible {P}rocesses}}
  (\bibinfo{publisher}{Interscience Publishers, Inc.}, \bibinfo{address}{New
  York}, \bibinfo{year}{1951}).

\bibitem[{\citenamefont{Qian and Hopfield}(1996)}]{Qian:1996uq}
\bibinfo{author}{\bibfnamefont{H.}~\bibnamefont{Qian}} \bibnamefont{and}
  \bibinfo{author}{\bibfnamefont{J.~J.} \bibnamefont{Hopfield}},
  \bibinfo{journal}{The Journal of Chemical Physics}
  \textbf{\bibinfo{volume}{105}}, \bibinfo{pages}{9292} (\bibinfo{year}{1996}).

\bibitem[{\citenamefont{Qian}(1998)}]{Qian:1998kx}
\bibinfo{author}{\bibfnamefont{H.}~\bibnamefont{Qian}}, \bibinfo{journal}{The
  Journal of Chemical Physics} \textbf{\bibinfo{volume}{109}},
  \bibinfo{pages}{10015} (\bibinfo{year}{1998}).

\end{thebibliography}

\pagebreak

\appendix

\section{Rate of change of the thermodynamic state variables}
\label{DotUSF}

After differentiating Eqns. (\ref{IntEnerSF}), (\ref{EntropySF}), and (\ref{FreeEnerSF}) we obtain the following results for the time derivatives of the internal energy:
\begin{eqnarray}
\dot{U} & = & \sum_{x} \dot{p}(x) (u_S(x)+ u_F(x)) + \sum_{x} p(x) \dot{u}_F(x), \nonumber \\
 & = & \displaystyle - \frac{k_B T}{2} \sum_{x,x'} (p(x') \Upsilon(x';x) - p(x) \Upsilon(x;x'))\log \frac{p^s(x)}{p^s(x')}  \nonumber \\
 & & \displaystyle - \frac{k_B T}{2} \sum_x p(x) \sum_{y,y'} (p(y'|x) \upsilon(x,y';x,y) - p(y|x) \upsilon(x,y;x,y')) \log \frac{p^s(y|x)}{p^s(y'|x)}, \nonumber
\end{eqnarray}
of the free energy:
\begin{eqnarray}
\dot{F} & = & \sum_x \dot{p}(x) \left( k_B T \log \frac{p(x)}{p^s(x)} + f_F(x) \right) + \sum_x p(x) \dot{f}_F(x), \nonumber \\
 & = & - \frac{k_B T}{2} \sum_{x,x'} (p(x') \Upsilon(x';x) - p(x) \Upsilon(x;x')) \log \frac{p(x')p^s(x)}{p(x)p^s(x')} \nonumber \\
 & & - \frac{k_B T}{2} \sum_x p(x) \sum_{y,y'} (p(y'|x) \upsilon(x,y';x,y) - p(y|x) \upsilon(x,y;x,y')) \nonumber \\
 & & \times \log \frac{p(y'|x) p^s(y|x)}{p(y|x) p^s(y'|x)}, \nonumber
\end{eqnarray}
and of the entropy:
\begin{eqnarray}
\dot{S} & = & \sum_x \dot{p}(x) \left( k_B \log p(x) + s_F(x) \right) + \sum_x p(x) \dot{s}_F(x), \nonumber\\
& = & \frac{k_B}{2} \sum_{x,x'} (p(x') \Upsilon(x';x) - p(x) \Upsilon(x;x')) \nonumber \\
& & \times \left( \log \frac{p(x') \Upsilon(x';x)}{p(x) \Upsilon(x;x')} - \log \frac{\Upsilon(x';x)}{\Upsilon(x;x')} \right) \nonumber \\
 & & + \frac{k_B T}{2} \sum_x p(x) \sum_{y,y'} (p(y'|x) \upsilon(x,y';x,y) - p(y|x) \upsilon(x,y;x,y')) \nonumber \\
 & & \displaystyle \times \left( \log \frac{p(y'|x) \upsilon(x,y';x,y)}{p(y|x) \upsilon(x,y;x,y')} -\log \frac{\upsilon(x,y';x,y)}{\upsilon(x,y;x,y')} \right). \nonumber
\end{eqnarray}
In the derivation of the previous equations we have taken into consideration that 
\[
\sum_x \dot{p}(x) u_F(x), \sum_x \dot{p}(x) f_F(x), \sum_x \dot{p}(x) s_F(x) \approx 0.
\]
The demonstration of these last relations is straightforward and follows from the fact that $\nu(x,y,x',y') \approx 0$ for all $x \neq x'$, which is the basic assumption underlying the time scale separation.

\end{document}